\documentclass{svjour3}

\usepackage[dvips]{graphicx}
\usepackage{booktabs}
\usepackage[courier]{altfont}
\usepackage{subfigure}
\usepackage{multirow}
\usepackage{listings}
\usepackage{graphicx}
\usepackage{color}
\usepackage{amsmath}
\usepackage{caption}
\usepackage{amssymb}
\usepackage{ulem}
\usepackage{url}
\usepackage{hyperref}
\usepackage{enumerate}

\begin{document}

\title{Prompt Learning for Multi-Label Code Smell Detection: A Promising Approach
}


\author{Haiyang~Liu$^{\ddag}$ \and Yang~Zhang$^{*,\ddag}$  \and Vidya~Saikrishna \and Quanquan~Tian \and Kun~Zheng 
    \thanks{$^{\ddag}$Co-first authors who contribute equally.} 
    \thanks{$^{*}$Corresponding author: uzhangyang@foxmail.com (Yang Zhang).}      
}


\institute{Haiyang Liu\at
              School of Information Science and Engineering, Hebei University of Science and Technology, Shijiazhuang, Hebei, China\\
              \email{liu15133450012@163.com}           \\
            Yang Zhang\at
              School of Information Science and Engineering, Hebei University of Science and Technology, Shijiazhuang, Hebei, China; Hebei Technology Innovation Center of Intelligent IoT, Shijiazhuang, Hebei, China\\
              \email{uzhangyang@foxmail.com}           \\
            Vidya Saikrishna\at
              School of Information Science and Engineering, Hebei University of Science and Technology, Shijiazhuang, Hebei, China\\
              \email{v.saikrishna@federation.edu.au}           \\
            Quanquan Tian\at
              School of Information Science and Engineering, Hebei University of Science and Technology, Shijiazhuang, Hebei, China\\
              \email{tianqq1216911@163.com}           \\
            Kun Zheng\at
              School of Information Science and Engineering, Hebei University of Science and Technology, Shijiazhuang, Hebei, China\\
              \email{17292225@qq.com}           \\
}

\date{Received: date / Accepted: date}

\maketitle

\begin{abstract}
Code smells indicate the potential problems of software quality so that developers can identify refactoring opportunities by detecting code smells. State-of-the-art approaches leverage heuristics, machine learning, and deep learning to detect code smells. However, existing approaches have not fully explored the potential of large language models (LLMs). 
In this paper, we propose \textit{PromptSmell}, a novel approach based on prompt learning for detecting multi-label code smell. Firstly, code snippets are acquired by traversing abstract syntax trees. Combined code snippets with natural language prompts and mask tokens, \textit{PromptSmell} constructs the input of LLMs. Secondly, to detect multi-label code smell, we leverage a label combination approach by converting a multi-label problem into a multi-classification problem. A customized answer space is added to the word list of pre-trained language models, and the probability distribution of intermediate answers is obtained by predicting the words at the mask positions. Finally, the intermediate answers are mapped to the target class labels by a verbalizer as the final classification result. We evaluate the effectiveness of \textit{PromptSmell} by answering six research questions. The experimental results demonstrate that \textit{PromptSmell} obtains an improvement of 11.17\% in $precision_{w}$ and 7.4\% in $F1_{w}$ compared to existing approaches.

\end{abstract}

\section{Introduction}
Code smell indicates that the basic design principles of object-oriented programming have been violated during the software design process, resulting in potential problems or poor design in the source code. Although the code smell does not necessarily indicate that the code is wrong, it often indicates that it may need further review or improvement. Therefore, detecting code smell plays a vital role in guiding subsequent code refactoring. Identifying and resolving potential code quality issues can help improve the readability and maintainability of the code, reducing the risk of software failure and the cost of future maintenance. 

The state-of-the-art approaches for code smell detection mainly include the heuristic approaches, the machine learning approaches, and the deep learning approaches, etc. iPlasma \cite{marinescu2005iplasma}, PMD \cite{pmd}, and DECOR \cite{moha2009decor} are typical tools for the detection of code smell based on a heuristic. Most of the tools analyze the source code statically and then obtain the metrics of the source code, which are compared with a predefined threshold of metrics. The code smell is present if the threshold is satisfied; otherwise, it is not present. Different developers may have other definitions and judgments of code smell. Therefore, the heuristic-based approach is affected by the subjective factors of the tool developers. Heuristic-based detection appears to lag in the face of the widespread use of machine learning (ML) techniques. For example, Kaur et. al., \cite{kaur2017support} proposed SVMCSD based on polynomial kernel functions of support vector machines to detect code smells. Furthermore, to address the subjectivity of the tool, Guggulothu et. al., \cite{guggulothu2020code} proposed a classifier chain-based approach to detect multiple code smells by leveraging a dataset with 61 class-level and 82 method-level metrics. With the success of deep learning (DL) techniques driven by image identification and data mining, researchers in industry and academia have shown great enthusiasm for applying deep learning techniques to solve various software engineering tasks. For example, Ananta et. al., \cite{das2019detecting} presented the use of metrics such as weight method class(WMC), tight class cohesion(TCC), and cyclomatic complexity(CC) as inputs of deep learning models to achieve the detection of brain classes and brain approaches. Liu et. al., \cite{liu2019deep} suggested a novel approach for training neural network classifiers in the realm of software development. They proposed incorporating both metric information and textual data processed through $Word2vec$ \cite{mikolov2013efficient} as inputs. This dual-input strategy aimed to enhance the classifiers' ability to detect code smells, specifically targeting code smells like feature envy and long methods.  However, traditional models such as $Word2vec$ have difficulty adequately capturing the semantic information of long contexts to generate word vectors. Hence, their semantic comprehension ability is relatively weak.

With the advent of the large language model, the trend of fine-tuning large language models utilizing large-scale data has emerged in recent years. Significant performance improvements have been achieved in many downstream tasks by exploiting the rich knowledge in the large language model corpus. For example, Zhang et. al. \cite{Zhang2022code} proposed to fine-tune the $BERT$ model by using a large amount of code text information and combining it with a traditional deep learning model to improve the performance of code smell detection.

Although the existing approaches have made significant progress in detecting code smells, we still face with several challenges. Firstly, the existing detection approaches are all based on fine-tuning pre-trained language models (PLMs) built on large-scale datasets, and the fine-tuning of the models and the collection of the datasets consumes a lot of time and effort. Secondly, a large amount of code metric information is contained in existing publicly available datasets, and the subjective factors of the researchers can influence the selection of these metrics. Finally, the traditional approaches exploit the PLMs to obtain word vectors of the code text information without taking full advantage of the potential domain knowledge acquired in the PLMs.

To address the aforementioned challenges, we propose \textit{PromptSmell}, a novel method for detecting multi-label code smells via prompt learning. 
This approach extracts code snippets using AST traversal and applies multiple detection rules for labeling code smells. It converts the multi-label problem into a multi-classification one through a label combination approach, generating unique target class labels. The PLMs predict words at mask positions, mapping them to target class labels using a verbalizer for the final multi-classification result.
To comprehensively evaluate the performance of \textit{PromptSmell} under different conditions, we conduct experiments by answering 6 research questions. 
The results demonstrate that \textit{PromptSmell} has a good performance in the four evaluation metrics of $accuracy$, $precision_{w}$, $recall_{w}$ and $F1_{w}$. 
Compared to fine-tuning, \textit{PromptSmell} demonstrates an improvement of 11.17\%  in precision and 7.4\% in F1 score, proving that the introduction of the prompt learning paradigm can make the model converge faster in small sample scenarios. 
Based on the above experimental results, we can conclude that introducing a prompt learning paradigm based on PLMs can effectively detect code smell, providing a practical cornerstone for future research.

The main contributions of this paper are summarized as follows.

\begin{enumerate}[(1)]
    \item We propose \textit{PromptSmell}, a novel approach to detect multi-label code smell based on prompt learning. To the best of our knowledge, this is the first work to utilize prompt learning to address the problem of code smell detection.

    \item \textit{PromptSmell} automatically extracts the source code from Java projects and builds a multi-label dataset containing only the source code of 19 real-world projects, which has been made publicly available at \href{https://github.com/HyLiu-cn/PromptSmell}{https://github.com/HyLiu-cn/PromptSmell}.

    \item We explore the scenario of data scarcity by applying the prompt learning paradigm in the domain of code smells, which provides a novel attempt to detect code smells.
\end{enumerate}

This paper describes the related work in Section \ref{2}. Section \ref{3} presents our motivation in detail, and a simple experiment is shown for proof and analysis. The few technical points involved in the \textit{PromptSmell} approach and the collection of the multi-label dataset are described in Section \ref{4}. The experimental setting, the research questions, and the evaluation metrics are presented in Section \ref{5}. Section \ref{6} describes several situations threatening this experiment. Finally, the conclusions of our experiment are drawn in Section \ref{7}, and future work is discussed.

\section{Related Work}
\label{2}
Existing code smell detection approaches are mainly categorized into heuristic-based detection, ML-based detection, and DL-based detection.

\subsection{Heuristic-based approach}
A heuristic-based approach is a common approach for code smell detection, and it depends on manually pre-designed sets of detection rules for each code smell. Terra et. al., \cite{terra2018jmove} proposed a JMove detection approach comparing the similarity of dependencies, which outperforms the current state-of-the-art JDeodorant tool. Tiwari et. al., \cite{tiwari2020functionality} presented a function-based detection approach for the long method code smell, which uses dependencies between the identified opportunities of different extraction methods as a rule to measure the code smell and achieved better detection results on Java open source code. Boutaib et. al., \cite{boutaib2021possibilistic} proposed ADIPOK-UMT a new search-based software engineering approach to address the problem of uncertainty in the thresholds of code metrics. The experimental results outperforms the best baseline approach. Besides the above, there are some typical heuristic-based detection approaches such as \cite{marinescu2005iplasma}, \cite{moha2009decor}, \cite{palomba2014mining}, etc., which can effectively help developers to identify and improve problems in their code. Although heuristic-based approaches are very effective in some cases, they lack consistency in the rules for detecting the same code smell and are susceptible to developer subjectivity.

\subsection{ML-based approach}
Many researchers have employed ML techniques in code smell detection to overcome the influence of subjective factors. Sandouka et. al., \cite{sandouka2023python} built a dataset of Python code smell containing 18 metric features and evaluated the long method code smell using six ML models. The results showed that random forest has the best performance in Python code smell detection. Sukkasem et. al., \cite{sukkasem2023enhanced} proposed an approach to improve the performance of ML-based detection by applying hyper-parameter optimization techniques in particle swarm optimization. The results of the experiments confirmed that the optimized ML classifiers outperformed the original version for code smell detection. Khleel et. al., \cite{khleel2023detection} proposed to evaluate a dataset with a random oversampling technique using five ML algorithms, and the result indicated that there is great promise in using ML techniques to predict code smell accurately. Zhang et. al., \cite{zhang2023code} proposed a new detection approach applying the stacking model and evaluated it in a large real-world project. The results demonstrated that the accuracy of code smell detection improved compared to existing approaches. Furthermore, the existing works such as \cite{brdar2022semi}, \cite{vatanapakorn2022python} have improved the level of detection by employing machine learning.

\subsection{DL-based approach}
With the advancement in the field of Deep Learning (DL), many researchers are adopting DL techniques to address the problem of code smell detection. Li et. al., \cite{li2022hybrid} proposed a hybrid model with multi-layer code representation to improve the code smell detection further. The result showed that the model was better than the existing detection approaches in multi-label code smell detection. Zhang et. al., \cite{zhang2022delesmell} expanded the positive samples in the dataset using refactoring tools to replace SMOTE and proposed a new GRU-CNN-based approach to detect code smells. The result of the experiment indicated that the approach improved the accuracy of brain class and brain method detection compared to the existing approaches. Bhave et. al., \cite{bhave2022deep} presented a multi-modal DL approach that combined structural and semantic information, and the experimental evaluation confirmed that the approach was better than the state-of-the-art techniques. In addition to the aforementioned approaches, a few other approaches based on deep learning such as \cite{zhang2021mars}, \cite{fang2020functional}, \cite{wang2020detection}, etc., provide novel ideas to detect code smell for more researchers. Although these approaches have achieved some results in code smell detection, the existing research mainly focuses on the detection of a single code smell and has not received sufficient attention for the case where multiple code smell exists in the code snippet. Meanwhile, which still needs to be further deeply researched and explored to improve the applicability and accuracy of the detection approaches in the real software engineering.

\section{Motivation}
\label{3}
This section leverages a motivating example to illustrate the problem of over-reliance on metrics information in existing detection approaches.

\begin{figure}[!htbp]
    \centering
    \includegraphics[scale=0.5]{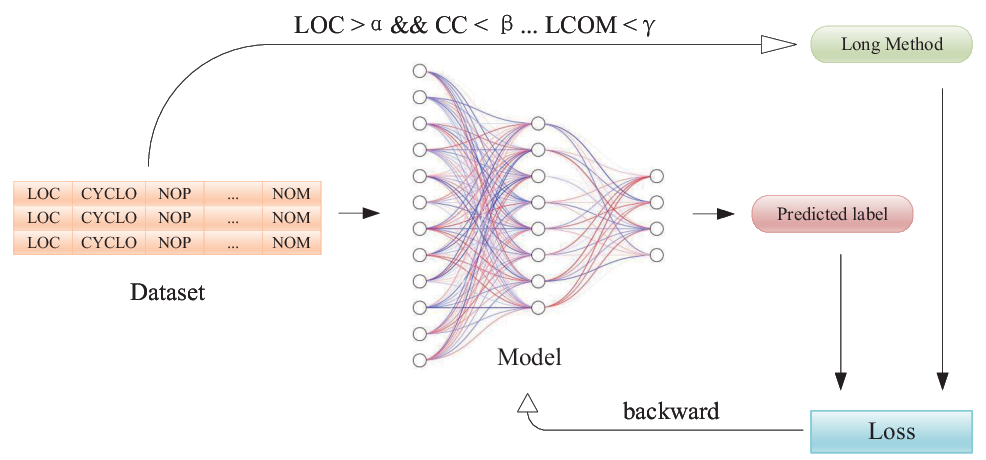}
    \caption{The training process of a model}
    \label{motivation1}
\vspace{-0.5cm}
\end{figure}

Existing datasets employed in many detection techniques include a variety of structural metrics. However, the essence of code smell labeling using both existing detection tools and detection rules is to set a threshold for one or more metrics, i.e., it is determined when a selected metric exceeds the set threshold as containing the code smell. Thus, the different structural metrics  used as inputs to the model,  captures the features of the input data and the weights are updated through backpropagation, as shown in Fig. \ref{motivation1}. However, the process may over-focus the model on a few specific metrics, resulting in the model's overfitting.

\begin{figure}[!htbp]
\vspace{-0.3cm}
    \centering
    \includegraphics[scale=0.28]{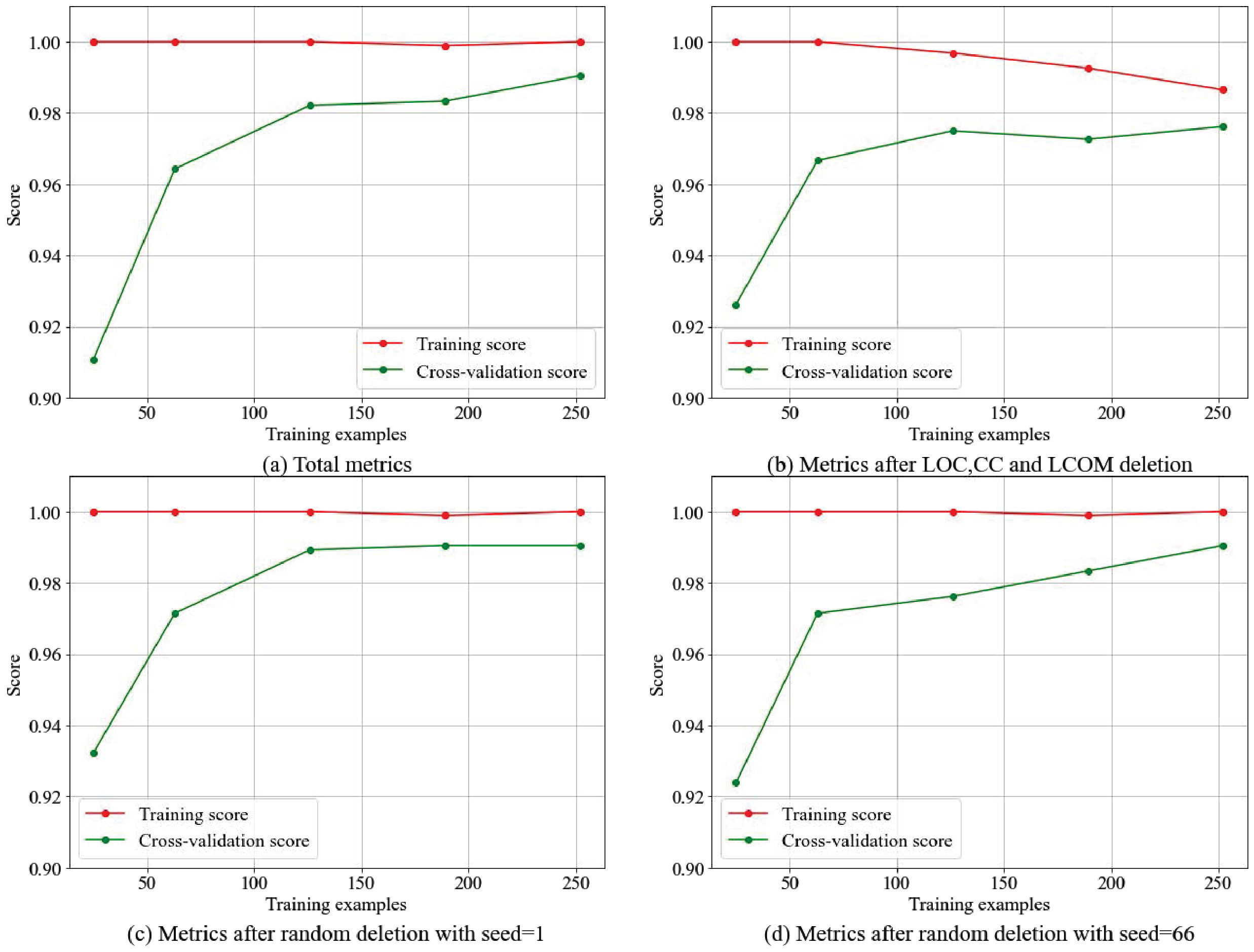}
    \caption{Results of comparison}
    \label{motivation2}
\vspace{-0.5cm}
\end{figure}

To demonstrate the aforementioned issues, we build a decision tree model \cite{safavian1991survey} and conduct experiments based on a publicly available dataset \cite{arcelli2016comparing}. The experimental results are shown in Fig.\ref{motivation2}, where Fig.\ref{motivation2}(a) represents the use of all metric data as inputs. Fig. \ref{motivation2}(b) represents the use of metric data with lines of code(LOC), cyclomatic complexity(CC), and lack of cohesion in methods(LCOM) deleted as inputs, Fig. \ref{motivation2}(c) represents the use of data following random deletion of the three metrics as inputs when the random seed is 1, Fig. \ref{motivation2}(d) represents the input using randomly deleting three metrics when the random seed is 66. 
The red curve represents the accuracy of the training set, and the green curve represents the accuracy of the test set. It can be observed from Fig. \ref{motivation2}(a) that the accuracy of the training set has been in the position of 1, while the accuracy of the test set has been gradually increasing from about 0.9 at the beginning, and the difference may be the result of overfitting of the model. The accuracy of the training set gradually decreases and converges to the accuracy of the test set in Fig. \ref{motivation2}(b). It indicates that the model gradually learns more general features during the training process by removing these metrics, which avoids overfitting and implies that the model is overly dependent on these metrics of the original dataset. Furthermore, we adopt the experiment of randomly removing the three metrics from the dataset to prove the above conclusions. It can be shown that the results of model training are not affected by randomly deleting the metrics in Fig. \ref{motivation2}(c) and \ref{motivation2}(d), and overfitting still exists. 

\section{The Approach}
\label{4}
This section presents an overview of \textit{PromptSmell}. Firstly, the general framework of \textit{PromptSmell} is presented. Then, we describe the essential steps in the data collection process. Finally, we provide a detailed description of the \textit{PromptSmell} project for building multi-labeled code smell.

\subsection{Overview}
\begin{figure}[!htbp]
    \centering
    \includegraphics[scale=0.38]{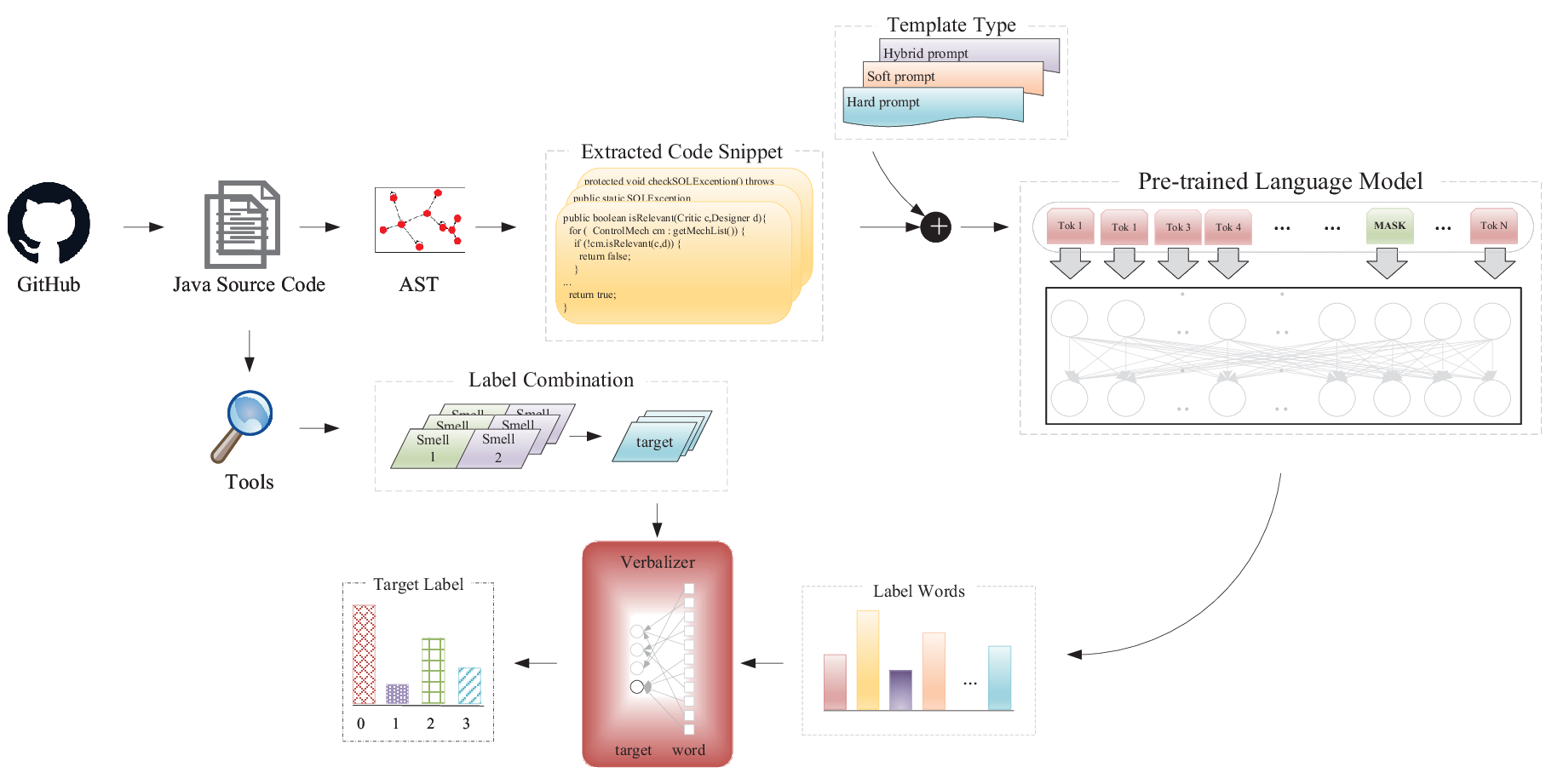}
    \caption{Overview of PromptSmell}
    \label{overview}
\vspace{-0.5cm}
\end{figure}

The overview of \textit{PromptSmell} is shown in Fig. \ref{overview}. Firstly, we select multi-domain Java projects from GitHub as the corpus to generate the dataset. Secondly, the extraction of the source code is completed using the iterative AST approach, and the code smell instances are labeled with multiple detection rules. The label combination approach is applied to convert the multi-label problem into a multi-classification problem to generate unique target class labels. The new data samples are generated by combining the extracted code snippets with the target class labels. Then, the self-defined answer spaces are added to the word list of the PLMs, and the new input text is generated by combining the code snippets with the prompt templates containing the natural language prompts and the mask token. The new input text is tokenized and used as an input to the PLMs, which is allowed to predict words at the mask position and, hence, obtain the probability distribution of the token words in the answer space at the mask position. Finally, we employ a verbalizer to map the token words predicted by the PLMs to the target class labels and obtain the probability distribution of the target class labels, using this as the final multi-classification prediction result.

\subsection{Dataset}
To ensure the diversity of the dataset, a large corpus of real Java projects from multiple domains is collected from GitHub as the dataset. The details of these projects are shown in Table \ref{datasets1}, where the first column is the name of the project, the second column is the description of the project, and the third, fourth, and fifth columns are the number of classes, the number of methods, and the number of lines of code, respectively. A total of 15143 classes are included in these 19 projects, with 52.6\% of the projects containing more than 500 classes. There are 139,446 methods involved, and the number of methods containing less than 1,000 is only 2 projects. The total lines of code is 1,620,112, with only 1 project containing less than 10,000 lines. 

\begin{table}[!h]
	\centering
    \renewcommand\arraystretch{0.8}
    \renewcommand{\tablename}{Table}
	\caption{Projects and their configurations}
	\scalebox{0.75}{
    \begin{tabular}{lp{8cm}lll}
		\hline
        Projects & Description & NOC & NOM & LOC
            \\ \hline
        ArgoUML  & A software application for drawing UML diagrams. & 1931 & 17129 & 157611
            \\
        Art of Illusion   & An open-source 3D modelling and rendering software in Java. & 487 & 6672 & 102679
            \\
        Cayenne & An open-source software that provides object-relational mapping and teleprocessing services. & 2983 & 17415 & 137928
            \\
        Cobertura & A tool for Java code coverage reporting. & 165 & 1156 & 14723
            \\
        Freecs & A testing program for an open-source free chat server. & 139 & 1404 & 20720
            \\
        Freemind & A software for mind mapping based on Java. & 532 & 7303 & 65866
            \\
        HSQLDB & A system of relational database management. & 664 & 13060 & 225106
            \\
        iText7 & An application for creating, adjusting, checking, and maintaining PDF documents. & 1624 & 12551 & 118717
            \\
        Ant-Ivy & A tool for managing (recording, tracking, resolving, and reporting) project dependencies. & 439 & 3087 & 21781
            \\
        Javacc & An application that generates syntax and lexical analyzers. & 180 & 1487 & 20886
            \\
        JHotDraw & A graphical editor framework that supports development in Java. & 488 & 4434 & 41105
            \\
        Maven & A tool for project construction, dependency management, and project information management of Java projects. & 731 & 3929 & 21703
            \\
        Mybatis & A tool to support custom SQL, stored procedures, and advanced mapping. & 118 & 534 & 3702
            \\
        Mylyn & A plugin for seamless task and context management integration into Eclipse. & 183 & 1064 & 10660
            \\
        ParallelColt & A high-performance computing tool that includes algorithms for data analysis, Fourier transforms, etc. & 1130 & 14527 & 216685
            \\
        Rhino & An open source project for executing JavaScript programs in the Java environment. & 628 & 6414 & 75341
            \\
        SPECjbb2005 & A benchmarking program for evaluating the performance of server-side Java applications. & 76 & 747 & 12716
            \\
        Xalan Java & A tool for converting XML documents to HTML, text, or other XML document types. & 963 & 10286 & 185511
            \\
        XML Graphics Batik & A tool for working with images in scalable vector graphics format. & 1682 & 16247 & 166672
            \\ \hline
        \textbf{Total}  &  & \textbf{15143} & \textbf{139446} & \textbf{1620112}
            \\ \hline
	\end{tabular}}
	\label{datasets1}
\vspace{-0.5cm}
\end{table}

To validate the performance of \textit{PromptSmell} in the scenario with scarce data, six projects are selected to generate a small sample dataset, including Drjava, Filecrush, Freeplane, JGroups, Nutch, and PMD. The descriptions of the six projects and their configurations are shown in Table \ref{datasets2}.

\begin{table}[!h]
	\centering
    \renewcommand\arraystretch{0.8}
    \renewcommand{\tablename}{Table}
	\caption{Projects and their configurations for generating training datasets of small samples}
	\scalebox{0.85}{
    \begin{tabular}{lp{8cm}lll}
		\hline
        Projects & Description & NOC & NOM & LOC
            \\ \hline
        Drjava & An open-source software for evaluating the capabilities of Java code. & 1145 & 16919 & 147245
            \\
        Filecrush & A utility program that fixes small files by merging them into larger files. & 23 & 311 & 6130
            \\
        Freeplane & A tool that provides mind mapping and navigation to map information. & 1112 & 9946 & 70019
            \\
        JGroups & A tool is a reliable group communication toolkit written entirely in Java. & 249 & 2005 & 13624
            \\
        Nutch & An open source framework for crawling the web content. & 370 & 2008 & 23298
            \\
        PMD & An Extensible Multilingual Static Code Analyser. & 2339 & 11063 & 56510
            \\ \hline

       \textbf{Total} & ~ & \textbf{5238} & \textbf{42252} & \textbf{316826}
            \\ \hline
	\end{tabular}}
	\label{datasets2}
\vspace{-0.5cm}
\end{table}

The dataset of \textit{PromptSmell} contains only code snippets of methods. The dataset is generated by an automated extraction tool \cite{zhang2022delesmell} to extract the method body of the source code. Firstly, the source code file is parsed into an abstract syntax tree, and the obtained object is stored in an ICompilationUnit object. Then, a subclass extending the ASTVisitor class implements the iteration of the abstract syntax tree nodes and saves the method body of the source code by accessing the MethodDeclaration type.

We select two types of code smell in this dataset: $Long\ Parameter\ List$ and $Long\ Method$. The selection of both code smells is based on a survey of code smell \cite{zhang2024code}. Furthermore, to ensure that instances of code smell are accurately labeled, we adopt multiple detection strategies for each type of code smell. The descriptions of the two code smells and the detection tools and rules applied are presented in Table \ref{rule}. The code snippet is labeled with the code smell if it satisfies both of our chosen detection tools and rules. Otherwise, it is labeled without the code smell.

\begin{table}[!htbp]
	\centering
    \renewcommand\arraystretch{1}
    \renewcommand{\tablename}{Table}
	\caption{Code Smell and Rules of Detection}
	\scalebox{0.78}{
	\begin{tabular}{lp{7.5cm}p{3.8cm}}
		\hline
        Code Smell & Description & Tools and Rules
        \\ \hline
        Long Parameter List & The code smell is present when a method has a long parameter list. & Designite\cite{Designite}, Danphitsanuphan rule\cite{danphitsanuphan2012code}
        \\
        Long Method & The code smell is present when the line of code of a method is excessively long. & Designite, Marinescu rule\cite{marinescu2005measurement}
        \\ \hline
	\end{tabular}}
	\label{rule}
\vspace{-0.5cm}
\end{table}

In this study, we adopt a label combination approach \cite{zhang2013review} to pre-process the multi-label dataset. The approach is to generate a new class for each combination between multiple labels and then convert the multi-label problem into a single-label multi-classification problem. For example, the multi-label code smell of a sample is $(TRUE, FALSE)$, which means that there is a $Long\ Parameter\ List$ and no $Long\ Method$ code smell, and a new unique label "1" is generated by applying the label combination approach. The reason for selecting the approach is that it captures the label dependencies (correlation or co-occurrence) during the classification process, which improves the classification performance \cite{guo2011multi}. Furthermore, the comparison between the multi-label code smell and the combination of labels to generate a new label is shown in Table \ref{mapping}.

\begin{table}[!htbp]
\vspace{-0.3cm}
	\centering
    \renewcommand\arraystretch{1}
    \setlength{\tabcolsep}{15pt}
    \renewcommand{\tablename}{Table}
	\caption{Mapping multiple labels into one representation}
	\scalebox{1}{
	\begin{tabular}{ccc}
		\hline
        \multicolumn{2}{c}{Multi Label}& \multirow{2}{*}{Label Combination} 
        \\
        Long Parameter List & Long Method & ~
        \\ \hline
        FALSE & FALSE & 0
        \\
        TRUE & FALSE & 1
        \\ 
        FALSE & TRUE & 2
        \\
        TRUE & TRUE & 3
        \\ \hline
	\end{tabular}}
	\label{mapping}
\vspace{-0.5cm}
\end{table}

\subsection{Prompt Engineering}
Prompt Template Engineering and Prompt Answer Engineering are the most critical parts of prompt engineering. The next step is to describe in detail how prompt template engineering and prompt answer engineering are designed.

\subsubsection{Prompt Template Engineering}
The prompt template engineering is the process of creating the prompt function $f_{prompt}=(X)$. The prompt function $f_{prompt}$ is used to "refactor" the input $X$ into $X^{'}$ \cite{10147172} for the most efficient performance of the downstream task.

To be closer to the pre-trained task of PLMs and to better motivate the performance of the PLMs \cite{liu2023pre}, the prompt templates are built using the $cloze$ format. In the current stage, the first step in building a prompt template is to consider the form of the template, which mainly contains $prefix$ prompts and $cloze$ prompts. An answer is generated when it is located at the end of the template, which is called $prefix$ prompt (e.g., "This method has 10 parameters. What is the code smell of this method? [Z]", where [Z] is the answer generated by PLMs Prediction). A position dug out of the middle of the template for answer generation is called $cloze$ prompt (e.g., "This method contains 10 parameters. The method has [Z] code smell.").

As this study is an early attempt, to provide more explicit guidance and hence more exactly direct the PLMs to generate output that matches the expectations, we select a manual creation approach, i.e., hard templates, based on domain knowledge. For example, the prompt function $f_{prompt}$ in the code smell classification task could be defined as:

\begin{equation}
    f_{prompt} = "The\ method\ has\ [Z]\ code\ smell.\ [X]"
    \label{fprompt}
\end{equation}

Various prompt functions can be obtained by replacing the natural language identifiers or mask positions. The prompt template is built with $f_{prompt}$ as shown in Table \ref{example}, where the input slot [X] is filled with code snippets and the output slot [Z] is used to generate an intermediate answer.

\begin{table}[!htbp]
	\centering
    \renewcommand\arraystretch{1}
    \renewcommand{\tablename}{Table}
	\caption{Example of the hard template}
	\scalebox{0.85}{
	\begin{tabular}{p{13.5cm}}
		\hline
        The method has [Z] code smell. public static void deleteAndRelease(Object mo){ Class c=mo.getClass(); TestCase.assertNotNull("toString() corrupt in " + c,mo.toString()); Model.getUmlFactory().delete(mo); Model.getPump().flushModelEvents(); TestCase.assertTrue("Could not delete " + c,Model.getUmlFactory().isRemoved(mo)); }
        \\ \hline
	\end{tabular}}
	\label{example}
\vspace{-0.3cm}
\end{table}

The types of templates in the prompt learning paradigm are mainly classified as hard templates, soft templates, and hybrid templates. To discuss the effect of different templates on the experimental results, we compare the three forms of templates and analyze the experimental results in \ref{RQ6} to demonstrate the effectiveness of hard template-based \textit{PromptSmell}.

\subsubsection{Prompt Answer Engineering}
\label{Prompt Answer Engineering}
Answer engineering aims to design an efficient answer space and map each token in it to a specific label of the target class.

The answer space is used to constrain the output of the PLMs so that intermediate answers are obtained for a specific task. There are three main types of answer shapes: Tokens, Span, and Sentence. Answers in the form of Token or Span are widely applied in classification tasks \cite{yin2019benchmarking}, while Sentence is more commonly employed in language generation tasks \cite{radford2019language}. Hence, our approach defines an intermediate answer as a token in the PLMs vocabulary. 

To obtain intermediate answers for the code smell classification task, we add the words related to code smell to the PLMs vocabulary as an answer space to limit the PLMs output. For example, the $Long\ Parameter\ List$ is defined as "long parameter list" and no code smell is described as "no", and they are appended to the vocabulary. Next, the PLMs predict [Z] in "The method contains 10 parameters, it has [Z] code smell", where Z $\in$ \{"no", "long parameter list "\}. 

Each target class label can correspond to one or more label words, i.e., one type of code smell can correspond to more than one natural language vocabulary. For example, $Feature\ Envy$ can be labeled with feature envy or denoted by using the abbreviation FE. We define the token corresponding to the type of code smell in the answer space as shown in Table \ref{space}. As this study is the first attempt to employ prompt engineering in code smell detection, we only define the token in the answer space using the letter case distinction.

\begin{table}[!htbp]
	\centering
    \renewcommand\arraystretch{1}
    \setlength{\tabcolsep}{15pt}
    \renewcommand{\tablename}{Table}
	\caption{Definition of code smell answer space}
	\scalebox{1}{
	\begin{tabular}{ll}
		\hline
        Code Smell & Answer Space
        \\ \hline
        Long Parameter List & Long Parameter List, long parameter list, LPL, lpl
        \\
        Long Method & Long Method, long method, LM, lm
        \\ \hline
	\end{tabular}}
	\label{space}
\vspace{-0.3cm}
\end{table}

In this study, we focus only on the output probability of the PLMs for the label words defined on the answer space, using the label words with the highest probability of being mapped to the target class as the predicted output. Specifically, let the training set be $\mathcal{D}=\{(X_{i},L_{i})\}_{i=1}^{N}$, and let the answer space be $\Pi=\{\pi_{j}\}_{j=1}^{n}$. We provide each sample $(X, L)$ as an input to the PLMs to obtain the probability distribution about each token $\pi$ under $X_{i}$. We follow by using the $topk$ algorithm to filter each token $\pi$ of the $i_{th}$ sample and select the token $\pi$ with the highest probability by sorting. The index $\mathcal{I}$ of the token $\pi$ is obtained:

\begin{equation}
    \mathcal{I}=argmax(p(\pi_{k}\mid X_{i})):=\{k\mid (k\in n)\wedge (\forall t\in n):p(\pi_{t}\mid X_{i})\leq p(\pi_{k}\mid X_{i})\}
    \label{index}
\end{equation}
Finally, we map the index $\mathcal{I}$ to the label of the target class, completing the entire classification process.

To investigate the performance of different answer spaces, the two different verbalizers are built:

\begin{align}
    \begin{split}
        Verbalizer_{1} &= \left\{\begin{array}{lcl}
                 0 & : &["no"] \\
                 1 & : &["long\ parameter\ list"] \\
                 2 & : &["long\ method"] \\
                 3 & : &["long\ method\ and\ long\ parameter\ list"]
                \end{array}\right.
    \end{split}
    \\
    \begin{split}
        Verbalizer_{2} &= \left\{\begin{array}{lcl}
                 0 & : &["no", "not", "zero"] \\
                 1 & : &["long\ parameter\ list", "lpl"] \\
                 2 & : &["long\ method", "lm"] \\
                 3 & : &["long\ method\ and\ long\ parameter\ list", "two", "all"]
                \end{array}\right.
    \end{split}
    \label{verbalizer}
\end{align}
and the experimental results are analyzed in \ref{RQ6}.

\section{Evaluation}
\label{5}
We first present the experimental configuration as well as research questions and then analyze the experimental results.

\subsection{Setup}
All experiments are conducted on a Lenovo workstation with an NVIDIA GTX1080Ti GPU and 16GB main memory running 64-bit Windows 10 with Python 3.10, PyTorch 2.0.1, and OpenPrompt 1.0.1 installed.

For the configurations of the experimental parameter, the number of iterations is $5$, the optimizer is $Adam$, the learning rate is set to $0.0001$, and the batch size is $4$. The hyperparameter settings of all models are shown in Table \ref{hyper-parameter}.

\begin{table}[!htbp]
\vspace{-0.3cm}
	\centering
    \renewcommand\arraystretch{1}
    \renewcommand{\tablename}{Table}
	\caption{The hyperparameter description and configuration of the model}
	\scalebox{1}{
	\begin{tabular}{llllll}
		\hline
		\multicolumn{2}{l}{Model} & \multicolumn{2}{l}{Hyper-parameter} & \multicolumn{2}{l}{Value}                     
                        \\ \hline
		
		\multirow{4}{*}{TextCNN} & ~          & embedding\_dim & ~            & 300 & ~
            \\ 
        ~ & ~                                 & filter region size & ~       & \{3,4,5\} & ~
            \\ 
        ~ & ~                                 & pooling & ~                  & 1-max pooling & ~
            \\ 
        ~ & ~                                 & activation function & ~      & ReLU & ~
                        \\ \hline

		\multirow{3}{*}{TextRCNN} & ~         & embedding\_dim & ~             & 300 & ~
            \\ 
        ~ & ~                                 & hidden\_size & ~               & 64 & ~
            \\ 
        ~ & ~                                 & pooling & ~                   & 1-max pooling & ~
                        \\ \hline

        \multirow{3}{*}{BiLSTM-Attention} & ~ & embedding\_dim & ~      & 300 & ~
            \\ 
        ~ & ~                                 & hidden\_size & ~       & 64 & ~
            \\ 
        ~ & ~                                 & attention\_out\_features & ~ & 128 & ~
                        \\ \hline
                        
		\multirow{2}{*}{\textit{PromptSmell}} & ~    & max\_seq\_length & ~        & 512 & ~
            \\
        ~ & ~                                 & truncate\_method & ~     & tail & ~
                        \\ \hline
	\end{tabular}}
	\label{hyper-parameter}
\vspace{-0.3cm}
\end{table}

\subsection{Research Questions}
We evaluate the effectiveness of \textit{PromptSmell} by answering six research questions (RQs).
\begin{description}
    \item[RQ1] How effective is \textit{PromptSmell} in detecting code smells?
	\item[RQ2] How effective is \textit{PromptSmell} compared to the fine-tuning paradigm?
	\item[RQ3] How effective is \textit{PromptSmell} compared to state-of-the-art models including TextCNN, BiLSTM-Attention, and TextRCNN?
	\item[RQ4] How effective is UniXcoder compared with BERT and GraphCodeBERT?
	\item[RQ5] How applicable is \textit{PromptSmell} for datasets with a very small amount of samples?
	\item[RQ6] What is the impact of different prompt settings on the performance of PLMs?
\end{description}

\subsection{Evaluation Metrics}
To evaluate the effectiveness of the model, we adopt four evaluation metrics \cite{2015A}, including \textit{accuracy}, \textit{precision}, \textit{recall}, and \textit{F1}. Furthermore, to more comprehensively assess the performance of the model on various classes, reducing the effect of unbalanced sample distributions, we apply weighted averages to \textit{precision}, \textit{recall}, and \textit{F1}, respectively. These metrics are presented as follows
\begin{align}
    accuracy &= \frac{TP + TN}{TP + FP + TN + FN}
        \\ \nonumber
    precision_{w} &= \sum_{i=1}^{m} (precision_{i} \cdot \frac{sample_{i}}{{\textstyle \sum_{j=1}^{m}}sample_{j}}) \\ 
        &=  \sum_{i=1}^{m} (\frac{TP_{i}}{TP_{i} + FP_{i}} \cdot \frac{sample_{i}}{{\textstyle \sum_{j=1}^{m}}sample_{j}})
        \\\nonumber
    recall_{w} &= \sum_{i=1}^{m} (recall_{i} \cdot \frac{sample_{i}}{{\textstyle \sum_{j=1}^{m}}sample_{j}}) \\
        &= \sum_{i=1}^{m} (\frac{TP_{i}}{TP_{i} + FN_{i}} \cdot \frac{sample_{i}}{{\textstyle \sum_{j=1}^{m}}sample_{j}})
        \\ \nonumber
    F1_{w} &= \sum_{i=1}^{m} (F1_{i} \cdot \frac{sample_{i}}{{\textstyle \sum_{j=1}^{m}}sample_{j}}) \\
         &= \sum_{i=1}^{m} (\frac{2 \cdot precision_{i} \cdot recall_{i}}{precision_{i} + recall_{i}} \cdot \frac{sample_{i}}{{\textstyle \sum_{j=1}^{m}}sample_{j}})
\end{align}
where true positive (TP) refers to the number of positive samples correctly predicted as positive classes, while true negative (TN) refers to the number of negative samples correctly predicted as negative classes, false positive (FP) signifies the number of negative samples falsely predicted as positive classes, and false negative (FN) indicates the number of positive samples wrongly classified as negative samples. $m$ represents the number of labels, and $sample_{i}$ is the number of samples of the $i_{th}$ label.

\subsection{Experimental Results}
The results of the experiment are analyzed and discussed, and the six research questions are answered in this section.

\subsubsection{RQ1: How effective is PromptSmell in detecting code smells}
To answer RQ1, the \textit{PromptSmell} is evaluated on six real-world projects. The experimental results are presented in Table \ref{rq1}. We observe that the $accuracy$ of \textit{PromptSmell} is 95.01\% for project PMD. For other projects, the $accuracy$ ranges from 90.50\% to 94.11\%. The $accuracy$ of project Nutch is the lowest compared to the other projects. The possible reason is that the features of the samples in this project are not fully considered by \textit{PromptSmell}. For project Filecrush , the $recall_{w}$ is 93.73\% and the $precision_{w}$ achieves 98.07\%. Similarly, project JGroups has 92.04\% $recall_{w}$ and 97.58\% $precision_{w}$. The reason is probably the insufficient number of positive samples in the project. Furthermore, the average $accuracy$ and average $precision_{w}$ of \textit{PromptSmell}'s detection on multi-label code smells are 93.08\%, the average $precision_{w}$ is 97.48\%, and the average $F1_{w}$ is 94.82\%. The results show that \textit{PromptSmell} has a good capability of generalization using source code as input data. It has a high $precision_{w}$ along with a high $recall_{w}$, which indicates that our approach is effective in minimizing false positives and detecting most of the instances of code smells.

\begin{table}[!h]
	\centering
    \renewcommand\arraystretch{1}
    \setlength{\tabcolsep}{13pt}
    \renewcommand{\tablename}{Table}
	\caption{The results of \textit{PromptSmell} in real-world projects}
	\scalebox{1}{
	\begin{tabular}{lcccc}
		\hline
         Projects & $accuracy$ & $precision_{w}$ & $recall_{w}$ & $F1_{w}$
        \\ \hline
        Drjava & 93.08\% & 97.90\% & 93.08\% & 94.97\%
            \\
        Filecrush & 93.73\% & 98.07\% & 93.73\% & 95.85\%
            \\
        Freeplane & 94.11\% & 98.01\% & 94.11\% & 95.56\%
            \\
        JGroups & 92.04\% & 97.58\% & 92.04\% & 94.29\%
            \\
        Nutch & 90.50\% & 94.95\% & 90.50\% & 91.92\%
            \\
        PMD & 95.01\% & 98.38\% & 95.01\% & 96.32\%
            \\ \hline
        Average & \textbf{93.08\%} & \textbf{97.48\%} & \textbf{93.08\%} & \textbf{94.82\%}
        \\ \hline
	\end{tabular}}
	\label{rq1}
\vspace{-0.3cm}
\end{table}

The above results show that \textit{PromptSmell} can make good use of source code as a dataset for code smell detection. It has stable detection ability and generalization performance.

\subsubsection{RQ2: How effective is PromptSmell compared to the fine-tuning paradigm?}
To answer RQ2, the \textit{PromptSmell} is compared to a traditional approach based on the fine-tuning paradigm, proving that using a prompt learning paradigm is better when using only source code.

\begin{figure}[!h]
    \centering
    \includegraphics[scale=0.28]{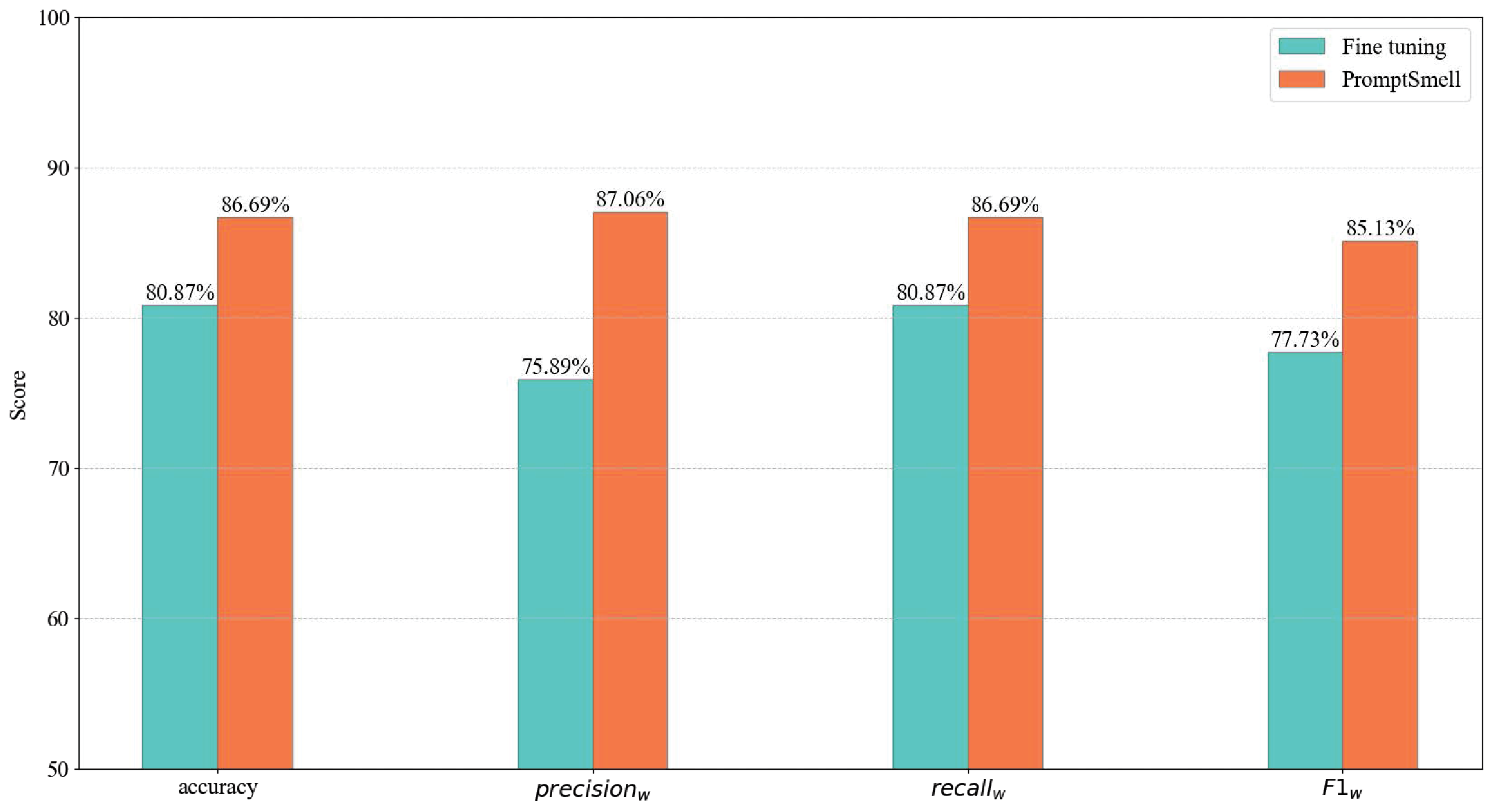}
    \caption{Comparison \textit{PromptSmell} against fine-tuning approach }
    \label{rq2}
\vspace{-0.3cm}
\end{figure}

The results of the two detection approaches are shown in Fig. \ref{rq2}. The figure shows that the $accuracy$ of \textit{PromptSmell} is 86.69\%, the $precision_{w}$ is 87.06\%, the $recall_{w}$ is 86.69\%, and the $F1_{w}$ is 85.13\%. The result indicates that the approach obtains a relatively excellent performance on the task of multi-label code smell classification. The largest improvement in $precision_{w}$ of \textit{PromptSmell} is 11.17\% (=87.06\%-75.89\%), and $F1_{w}$ is improved by 7.4\% (=85.13\%-77.73\%) when compared to an approach based on the fine-tuning paradigm. Moreover, both $accuracy$ and $recall_{w}$ metrics are increased from 80.87\% to 86.69\%. The results prove that PLMs has a better ability to extract semantic information in context, and \textit{PromptSmell} fully exploits the domain knowledge of PLMs's previous learning by introducing the prompt learning paradigm, leading to an improvement in the detection performance.

The above results show that \textit{PromptSmell} is more efficient and comprehensive in utilizing the existing knowledge of PLMs than an approach based on the fine-tuning paradigm and better performs detecting the multi-label code smell.

\subsubsection{RQ3: How effective is PromptSmell compared to state-of-the-art models including TextCNN, BiLSTM-Attention, and TextRCNN?}

Since only code snippets are employed in our dataset, to evaluate the effectiveness of considering code snippets as a text classification task, we select three classification models (TextCNN \cite{zhang2015sensitivity}, BiLSTM-Attention \cite{zhou2016attention}, and TextRCNN \cite{lai2015recurrent}) to compare with \textit{PromptSmell}. The performance of all models is presented in Table \ref{rq3}.

It is observed that the performance of the TextCNN model is poor compared to \textit{PromptSmell} from Table \ref{rq3}. Specifically, the checking $accuracy$ of TextCNN is only 41.09\%, which is 45.97\% (=87.06\%-41.09\%) less compared to \textit{PromptSmell}'s 87.06\%, and $F1_{w}$ is 37.49\% (=85.13\%-47.64\%) lower. The result indicates that \textit{PromptSmell} has a more powerful ability to extract features on this task, resulting in a higher performance. For the BiLSTM-Attention model, there is a small performance improvement compared to TextCNN. However, the BiLSTM-Attention has a large gap compared to \textit{PromptSmell}. Specifically, the $accuracy$ of \textit{PromptSmell} is increased from 60.64\% to 86.69\%, the $precision_{w}$ is improved from 55.77\% to 87.06\%, and the $F1_{w}$ is improved from 53.30\% to 85.13\%. Furthermore, we observe that the recall of BiLSTM-Attention is approximately 5\% higher than the $precision_{w}$, indicating the performance of the model is probably unstable. The $accuracy$, $precision_{w}$, $recall_{w}$, and $F1_{w}$ of \textit{PromptSmell} are improved by 5.73\% (=86.69\%-80.96\%), 6.22\% (=87.06\%-80.84\%), 5.73\% (=86.69\%-80.96\%), and 5.47\% (=85.13\%- 79.66\%), respectively. The result fully demonstrates that \textit{PromptSmell} has a better ability to extract semantic information from the context by fully exploiting the existing natural language and programming language knowledge of PLMs.

\begin{table}[!htbp]
	\centering
    \renewcommand\arraystretch{1.2}
    \setlength{\tabcolsep}{13pt}
    \renewcommand{\tablename}{Table}
	\caption{Comparsion between PromptSmell and  existing models}
	\scalebox{0.9}{
	\begin{tabular}{lcccc}
		\hline
        Model & $accuracy$ & $precision_{w}$ & $recall_{w}$ & $F1_{w}$
        \\ \hline
        TextCNN & 58.64\% & 41.09\% & 58.64\% & 47.64\%
        \\
        BiLSTM-Attention & 60.64\% & 55.77\% & 60.64\% & 53.30\%
        \\
        TextRCNN & 80.96\% & 80.84\% & 80.96\% & 79.66\%
        \\
        \textit{PromptSmell} & \textbf{86.69\%} & \textbf{87.06\%} & \textbf{86.69\%} & \textbf{85.13\%}
        \\ \hline
	\end{tabular}}
	\label{rq3}
\vspace{-0.3cm}
\end{table}

The result of the experiment shows that \textit{PromptSmell} has a better performance than existing models of text classification. By "refactoring" the input data and adding prompts that are relevant to the downstream task, \textit{PromptSmell} utilizes the domain knowledge of PLMs more comprehensively and efficiently, which improves the performance of detection.

\subsubsection{RQ4: How effective is UniXcoder compared with BERT and GraphCodeBERT?}
To answer RQ4, we compare the UniXcoder\cite{guo2022unixcoder} used in \textit{PromptSmell} with BERT \cite{devlin2018bert} and GraphCodeBERT \cite{guo2020graphcodebert}. The experimental results are shown in Table \ref{rq4}.

\begin{table}[!htbp]
	\centering
    \renewcommand\arraystretch{1.2}
    \setlength{\tabcolsep}{15pt}
    \renewcommand{\tablename}{Table}
	\caption{Comparison with other PLMs}
	\scalebox{1}{
	\begin{tabular}{lccc}
		\hline
        \multirow{2}{*}{} & \multicolumn{3}{c}{\textit{PromptSmell}}
        \\ \cline{2-4}
        ~ & BERT & GraphCodeBERT & UniXcoder
        \\ \hline
        $accuray$ & 30.87\% & 83.44\% & \textbf{86.69\%}
        \\
        $precision_{w}$ & 9.53\% & 83.46\% & \textbf{87.06\%} 
        \\
        $recall_{w}$ & 30.87\% & 83.44\% & \textbf{86.69\%} 
        \\
        $F1_{w}$ & 14.56\% & 82.30\% & \textbf{85.13\%} 
        \\ \hline
	\end{tabular}}
	\label{rq4}
\vspace{-0.3cm}
\end{table}

The results show that the $accuracy$ is 86.69\%, the $precision_{w}$ is 87.06\%, the $recall_{w}$ is 86.69\% and the $F1_{w}$ is 85.13\% when the UniXcoder model is used as the base PLMs for \textit{PromptSmell}. The worst performance of \textit{PromptSmell} is achieved when we change the UniXcoder model to the BERT model. It shows that the $accuracy$ and $recall_{w}$ are only 30.87\%, the $precision_{w}$ is only 9.53\% and the $F1_{w}$ is only 14.56\%. The reason for the large gap between the two is that the BERT model is a common PLM, it has a good performance for tasks of natural language processing. However, a bimodal language model like UniXcoder is more advantageous when dealing with specific domains such as programming languages. The PromptSmell demonstrates a relatively good performance when the base PLM is the GraphCodeBERT model. However, the $accuracy$ and $recall_{w}$ are 3.25\% (=86.69\%-83.44\%) lower than the original approach, the $precision_{w}$ is 3.6\% (=87.06\%-83.46\%) lower, and the $F1_{w}$ is 2.83\% (=85.13\%- 82.30\%). The difference shows that the UniXcoder model learns a better representation of task relevance than the GraphCodeBERT model on our code smell dataset, which improves the overall performance of the approach.

The results of the above comparison conclude that the performance of the UniXcoder model is better compared to the BERT model and the GraphCodeBERT model in our multi-label code smell dataset. It is implicitly proven that the UniXcoder model has more knowledge about the relevant domain. Therefore, it can understand the source code more correctly.

\subsubsection{RQ5: How applicable is PromptSmell for datasets with a very small amount of samples?}
To ensure the reliability of the experimental results, we regenerate the training set. Zero-sample and small-sample scenarios are modeled by controlling the total amount of data in the training set (0 samples, 64 samples, 256 samples, 512 samples, and 1024 samples) to evaluate the effectiveness of \textit{PromptSmell}. Furthermore, we ensure that the same data does not exist in each small-sample dataset by setting a random seed.

\begin{table}[!htbp]
	\centering
    \renewcommand\arraystretch{1.2}
    \setlength{\tabcolsep}{15pt}
    \renewcommand{\tablename}{Table}
	\caption{The results of \textit{PromptSmell}'s performance in the small sample}
	\scalebox{0.9}{
	\begin{tabular}{ccccc}
		\hline
       \# Training samples & $accuracy$ & $precision_{w}$ & $recall_{w}$ & $F1_{w}$
        \\ \hline
        0 & 31.58\% & 52.81\% & 31.58\% & 17.67\%
        \\
        64 & 49.52\% & 50.07\% & 49.52\% & 39.23\%
        \\
        256 & 57.11\% & 52.59\% &  57.11\% & 52.53\%
        \\
        512 & 74.00\% & 72.07\% & 74.00\% & 71.75\%
        \\
        1024 & 76.48\% & 73.42\% & 76.48\% & 74.06\%
        \\ \hline
	\end{tabular}}
	\label{rq5}
\vspace{-0.3cm}
\end{table}

The results of the \textit{PromptSmell} detection in randomly generated scenarios with zero or small samples are shown in Table \ref{rq5}. It is observed that the $accuracy$ of \textit{PromptSmell} is 31.58\%, the $precision_{w}$ is 52.81\%, and the $F1_{w}$ is only 17.67\% in the scenario with zero-sample scenario, indicating that the domain-specific knowledge in PLMs is not fully exploited without using training samples for training. The performance of \textit{PromptSmell} improves significantly when the training samples are only 64. Specifically, the $accuracy$ and $recall_{w}$ are increased by 17.94\% (=49.52\%-31.58\%), and the $F1_{w}$ is improved from 17.67\% to 39.23\%. The result indicates that the \textit{PromptSmell} is effective in dealing with data scarcity and achieves a significant improvement in performance by adding a small number of training samples. The $accuracy$, $precision_{w}$, $recall_{w}$, and $F1_{w}$ of \textit{PromptSmell} are 57.11\%, 52.59\%, 57.11\% and 52.23\% respectively when 256 samples are included in the training set. When compared to 64 training samples, the $precision_{w}$ is not improved significantly, while the $F1_{w}$ metric is improved by 13.3\% (=52.53\%-39.23\%), indicating that the generalization of the approach has been effectively improved. We observe that the approach improves by at least 16.89\% (= 74.00\% - 57.11\%) on all evaluation metrics when the number of training samples is increased to 512. Furthermore, we notice that the $accuracy$ of the approach increases from 74.00\% to 76.48\% and $F1_{w}$ from 71.75\% to 74.06\% as the amount of data in the training set is further increased to 1024, while the increase of $precision_{w}$ is slower, increasing by only 1.35\% (=73.42\%-72.07\%). The result demonstrates that \textit{PromptSmell} achieves faster convergence and better performance by training with less data. 

The above results conclude that \textit{PromptSemll} has faster convergence and better performance on small sample datasets. The performance advantage is more significant especially when the training data is extremely scarce.

\subsubsection{RQ6: What is the impact of different prompt settings on the performance of PLMs?}
\label{RQ6}
The format of the prompt template and the contents of the verbalizer are shown in Table \ref{rq6-1}. Three types of templates are built to investigate the effect of templates on classification results. P1 in the table shows the building of hard templates by manually adding natural language. For the building of soft templates, the natural language in the hard templates is replaced with soft tokens $[SOFT]$ as shown in P2 in the table. Finally, for the mixed templates, we preserve the keywords in the hard templates that capture the downstream tasks and replace the unimportant words with the same soft tokens in the soft templates as shown in P3 in the table. Furthermore, two verbalizers in \ref{Prompt Answer Engineering} are utilized to investigate the effect of different answer spaces.

\begin{table}[!htbp]
	\centering
    \renewcommand\arraystretch{1}
    \renewcommand{\tablename}{Table}
	\caption{The various prompt templates and label mapping words}
	\scalebox{0.79}{
	\begin{tabular}{lp{4.1cm}p{3.4cm}p{3.4cm}}
		\hline
        Prompt templates & P1: The method has [MASK] code smell. [CODE] & P2: [SOFT]*3 [MASK] [SOFT]*2. [CODE] & P3: [SOFT]*3 [MASK] code smell. [CODE]
        \\ \hline
        \multirow{10}{*}{Verbalizer} & \multicolumn{3}{l}{V1:} 
        \\
        ~ & \multicolumn{2}{l}{0:["no"]}
        \\
        ~ & \multicolumn{2}{l}{1:["Long Parameter List"]}
        \\
        ~ & \multicolumn{2}{l}{2:["Long Method"]}
        \\
        ~ & \multicolumn{2}{l}{3:["Long Method and Long Parameter List"]}
        \\ \cline{2-4}
        ~ & \multicolumn{2}{l}{V2:}
        \\
        ~ & \multicolumn{2}{l}{0:["no", "not", "zero"]}
        \\
        ~ & \multicolumn{2}{l}{1:["Long Parameter List", "lpl"]}
        \\
        ~ & \multicolumn{2}{l}{2:["Long Method", "lm"]}
        \\
        ~ & \multicolumn{2}{l}{3:["Long Method and Long Parameter List", "two", "all"]}
        
        \\ \hline
	\end{tabular}}
	\label{rq6-1}
\vspace{-0.3cm}
\end{table}

The UniXcoder model with PL and NL training is selected as the base PLMs for this experiment. To examine the effect of different prompt settings on the performance of the PLMs, the prompt templates in Table \ref{rq6-1} are randomly combined with the verbalizer.

The result of the random combination experiment is shown in Table \ref{rq6-2}. It is observed that the four evaluation metrics of \textit{PromptSmell} are the lowest ($accuracy$ is 61.02\%, $precision_{w}$ is 43.97\%, $recall_{w}$ is 61.02\% and $F1_{w}$ is 50.47\%) when the verbalizer is V1 with the hybrid template P3. However, we observe a significant improvement in each of the evaluation metrics when the verbalizer is V2. The maximum improvement in $precision_{w}$ is 41.98\% (=85.95\%-43.97\%). Furthermore, the phenomenon is repeated when using the other two types of templates. Specifically, it is shown that the $accuracy$ increases from 62.36\% to 86.69\%, the $precision_{w}$ improves from 45.43\% to 87.06\%, the $recall_{w}$ improves from 62.36\% to 86.69\% and the $F1_{w}$ increases from 51.69\% to 85.13\% when using V2 compared to V1 under P1. The $accuracy$ is 61.21\%, the $precision_{w}$ is 46.06\%, the $recall_{w}$ is 61.21\%, and the $F1_{w}$ is 51.26\% when using the V1 verbalizer under the soft template P2. However, the improvement in $precision_{w}$ is 39.99\% (=86.05\%-46.06\%) and $F1_{w}$ is 33.92\% (=85.18\%-51.26\%) using the V2 verbalizer. The above result indicates that the existing domain knowledge that the PLMs have acquired can be effectively motivated by the addition of label words with similar meanings to improve the overall performance of \textit{PromptSmell}. Furthermore, we also observe that \textit{PromptSmell} obtains the highest $accuracy$ (86.69\%), $precision_{w}$ (87.06\%) and $recall_{w}$ (86.69\%) in the combination of P1-V2. The best overall performance of all combinations is obtained by the P2-V2 combination, with the $F1_{w}$ is 85.18\%, which is higher than that of the 85.13\% for the P1-V2 combination. The result shows that soft templates may offer more flexibility than hard templates, resulting in a model better adapted to the variation and complexity of the task. However, it is undesirable to obtain a 0.05\%(=85.18\%-85.13\%) improvement in performance by sacrificing the efficiency of the model in our task.

\begin{table}[!htbp]
	\centering
    \renewcommand\arraystretch{1.2}
    \setlength{\tabcolsep}{12pt}
    \renewcommand{\tablename}{Table}
	\caption{The results of different templates and verbalizers}
	\scalebox{0.9}{
	\begin{tabular}{lcccccc}
		\hline
        ~ & P1-V1 & P1-V2 & P2-V1 & P2-V2 & P3-V1 & P3-V2 
        \\ \hline
        $accuray$ & 62.36\% & \textbf{86.69\%} & 61.21\% & 86.59\% & 61.02\% & 85.64\% 
        \\
        $precision_{w}$ & 45.43\% & \textbf{87.06\%} & 46.06\% & 86.05\% & 43.97\% & 85.95\%
        \\
        $recall_{w}$ & 62.36\% & \textbf{86.69\%} & 61.21\% & 86.59\% & 61.02\% & 85.64\% 
        \\
        $F1_{w}$ & 51.69\% & 85.13\% & 51.26\% & \textbf{85.18\%} & 50.47\% & 84.20\% 
        \\ \hline
	\end{tabular}}
	\label{rq6-2}
\vspace{-0.3cm}
\end{table}

The results show that the design of templates and verbalizers affects the performance of \textit{PromptSmell} in our multi-label code smell classification task. However, the design of the verbalizer has a more significant effect on \textit{PromptSmell} than the different templates, indicating that the performance of \textit{PromptSmell} can be significantly improved by the manual addition of label words with similar meanings.

\section{Threats to Validity}
\label{6}
This section discusses threats to validity that may affect the results of the experiment.

\subsection{Internal Validity}
Internal validity concerns the causal relationships within the experiment and the reasonableness of the experimental design. We employ UniXcoder as the PLMs in \textit{PromptSmell}. The UniXcoder is widely used for code-related understanding and generation tasks. Although the model performs better on the code smell types of our selection, it may lack the understanding of other kinds of code smell, which affects our results to a certain extent. Furthermore, the design of templates and verbalizers mainly affects the effectiveness of \textit{PromptSmell} in generating prompt and answer mapping. To mitigate this threat effectively, we have designed and optimized the templates and verbalizers to ensure their efficiency and adaptability in the code smell detection task. We have compared them with other types of templates or verbalizers to increase our confidence in the experimental results.

\subsection{External Validity}
External validity is concerned with the ability to generalize experimental results. We evaluate \textit{PromptSmell} on real Java projects in different domains, and further experiments should be performed on other projects to demonstrate that \textit{PromptSmell} has a good ability for generalization. Furthermore, due to differences in the distribution of code smell between languages, it is necessary to perform studies on the transfer of \textit{PromptSmell} to other languages.


\section{Conclusion}
\label{7}

We propose a multi-label code smell detection approach \textit{PromptSmell} based on prompt learning by applying a large model-based prompt learning paradigm. The approach builds on PLMs by using a prompt tuning approach with specific templates to convert the data of downstream tasks into a natural language form, which fully exploits the ability of the PLMs. 
Specifically, code snippets are obtained by AST and multiple detection rules are applied to label code smell instances. Then, a label combination approach is used to generate unique target class labels, and the code snippets are input into the PLM in combination with natural language prompts and special mask tokens. Finally, the answer space tokens predicted by the PLM are mapped to the actual target class labels, and the probability distribution of the target class labels is obtained, which is used as the final multi-classification prediction result. 
We evaluate the effectiveness of \textit{PromptSmell} by answering six research questions. The experimental results show that \textit{PromptSmell} has good performance in $accuracy$, $precision_{w}$, $recall_{w}$ and $F1_{w}$. The  $precision_{w}$ is improved by 11.17\% and $F1_{w}$ is improved by 7.4\% compared to the existing approaches. Based on the above experimental results, we can observe that introducing a prompt learning paradigm based on PLMs effectively detects code smell. In the future we will investigate the detection of other code smells to evaluate their extensibility and effectiveness.






\bibliographystyle{elsarticle-num}
\bibliography{zhangyangref}







\end{document}